\documentstyle[sprocl,epsfig]{article}

\def\be{\begin{equation}}
\def\ee{\end{equation}}
\def\bea{\begin{eqnarray}}
\def\eea{\end{eqnarray}}
\def\bq{\begin{eqnarray}}
\def\eq{\end{eqnarray}}
\begin{document}
\centerline{\footnotesize (Preprint: WUE-ITP-00-020, 
NIKHEF-00-021, UM-TH-00-20)}
\title{QCD SUM RULES FOR HEAVY FLAVORS
\footnote{Talk given by O.Yakovlev at the 4th Workshop on
Continuous Advances in QCD, Minneapolis, May 12-14, 2000 }}
\author{OLEG YAKOVLEV}
\address{Randall Laboratory of Physics, University of Michigan,\\ 
Ann Arbor, Michigan 48109-1120, USA}
\author{REINHOLD R\"UCKL}
\address{Institut f\"ur Theoretische Physik, Universit\"at W\"urzburg,\\
D-97074 W\"urzburg, Germany}
\author{STEFAN WEINZIERL}
\address{NIKHEF, P.O. Box 41882, NL - 1009 \\DB Amsterdam, The Netherlands}
\maketitle\abstracts{
We give a short review of QCD sum rule results for $B$ and $D$ 
mesons and $\Lambda_Q$ and $\Sigma_Q$ baryons. 
We focus mainly on recent developments concerning  
semileptonic $B\to\pi$ and $D\to\pi$ transitions, 
 pion couplings to heavy hadrons,  
  decay constants and estimates of the $b$ quark mass  
from a baryonic sum rule, and the extraction 
of the pion distribution amplitude from CLEO data.} 
\section{Introduction}
The accurate study of $B$ meson decays is a main source of information 
for understanding $CP$ violation and the physics of heavy quarks. 
In particular, experiments at $B$ factories will allow measurements 
of $B$ decay properties with good precision \cite{BaBar}.  
 On the theoretical side, the method of QCD sum rules\cite{SVZ} 
remains one of the main tools in applying Quantum Chromodynamics to  
hadron physics.   
Since its birth in 1979, the sum rule method has become more and more 
advanced not only technically, but also conceptually.  
In this talk, we give a short review of QCD sum rule results 
for $B$ and $D$ mesons and $\Lambda_Q$ and $\Sigma_Q$ baryons. 
 We focus mainly on recent developments concerning semileptonic 
$B\to\pi$ and $D\to\pi$ transitions\cite{K2000},
including a new approach\cite{WY}, which will be discussed in detail,    
and new results on the $f^0$ form factor \cite{Fzero}. 
Furthermore we will address pion couplings to $B$ and $D$ mesons 
and to $\Lambda_Q$ and  $\Sigma_Q$ baryons,   
meson decay constants and corresponding matrix element for baryons, 
an estimate of the $b$ quark mass from a baryonic sum rule, 
and finally a recent extraction of the pion distribution 
amplitude from CLEO data\cite{Schm}. 
\section{Pion distribution amplitude from CLEO data}
We start with the pion distribution amplitudes, which 
serve as input in the QCD sum rule method and allow 
the calculation of heavy-to-light form factors (e.g., 
 $f^+_{B\to\pi}$ and  $f^+_{D \to \pi}$)  
and  hadronic coupling constants (e.g.,  $g_{B^*B\pi}$ and  
$g_{D^*D\pi}$).  Recently, the CLEO collaboration has measured the 
$\gamma\gamma^*\to \pi^0$ form factor. 
In this experiment 
\cite{CLEO} \footnote{There also exist older results from  
the CELLO collaboration \cite{CELLO}.},  one of the photons 
is nearly on-shell and the other one is highly 
off-shell, with a virtuality in the range $1.5$ GeV$^2$ -- $9.2$ 
GeV$^2$.
The possibility of extracting the twist-2 pion 
distribution amplitude from the CLEO data has been 
studied in the papers \cite{Schm,AK}. 
There, the light-cone sum rule (LCSR) method has been used    
to calculate the relevant form factor and to compare the calculation 
with the measurement of $\gamma\gamma^*\to \pi^0$. 
 
In order to sketch the basic idea we begin with the correlator  
of two vector currents $j_{\mu}=(\frac{2}{3}\bar u \gamma_{\mu} u 
-\frac{1}{3}\bar d \gamma_{\mu} d )$:
\begin{eqnarray}\label{Correlator}
\int d^4 x e^{-iq_1x}\langle \pi^0(0)|T\{j_\mu(x) j_\nu(0) \}|0 \rangle =
i\epsilon_{\mu\nu\alpha\beta}q_1^\alpha q_2^\beta 
F^{\pi\gamma^*\gamma^*}(s_1,s_2),
\end{eqnarray}
where $q_1, q_2$ are the momenta of the photons, and  $s_1=q_1^2$,
 $s_2=q_2^2$ are the virtualities. 
 In the CLEO data, one of the virtualities is small, 
i.e. $s_2\to 0$.
Since a straightforward  OPE  calculation is impossible, 
 we have to use analyticity and duality arguments.
One can write the form factor as a dispersion relation in $s_2$:
\begin{equation}\label{E:hr}
F^{\pi\gamma^*\gamma^*}(s_1,s_2) = \frac{\sqrt{2}\,f_\rho\,F^{\rho\pi}(s_1)}
 {m_\rho^2-s_2} + \int\limits_{s_0}^\infty\!\!ds\,\frac{\rho^h(s_1,s)}
 {s-s_2}.
\end{equation}
For the physical ground states $\rho$ and $\omega$ we take 
 $m_\rho\simeq m_\omega$;   
$\frac{1}{3}\langle \pi^0(p)|j_\mu|\omega(q_2)\rangle \simeq
 \langle \pi^0(p)|j_\mu|\rho^0(q_2)\rangle = 
  \frac{1}{m_\rho}\,\epsilon_{\mu\nu\alpha\beta}\,e^\nu\,q_1^\alpha
  \,q_2^\beta\,F^{\rho\pi}(s_1)$;  
$3\,\langle\omega|j_\nu|0\rangle \simeq \langle\rho^0|j_\nu|0\rangle  =
  \frac{f_\rho}{\sqrt{2}}\,m_\rho\,e^*_\nu$, 
$e_{\nu}$ being  the polarization vector of the $\rho$ meson and  
$f_\rho$ being the decay constant. 
 The spectral density of the higher energy states $\rho^h(s_1,s)$ is derived 
from the expression for  $F_{QCD}^{\pi\gamma^*\gamma^*}(s_1,s)$
 calculated in QCD, assuming semi-local quark-hadron 
duality for  $s>s_0$.  
 Equating the dispersion relation (\ref{E:hr}) 
with the QCD expression at large  $s_2$, and 
performing a Borel transformation in $s_2$, 
one gets the LCSR: 
\begin{equation}\label{E:rho}
\sqrt{2}\,f_\rho\,F^{\rho\pi}(s_1) = \frac{1}{\pi}\,\int\limits_0^{s_0}\!\! ds\,
 \mbox{Im}\,F_{QCD}^{\pi\gamma^*\gamma^*}(s_1,s)\, 
  \mbox{e}^{\frac{m_\rho^2-s}{M^2}},
\end{equation}
where $M$ is the Borel parameter. 
Substituting  (\ref{E:rho}) into (\ref{E:hr}) and taking  
$s_2\to 0$ one finally obtains  
\begin{equation}\label{E:srggpi}
F^{\pi\gamma\gamma^*}(s_1) = \frac{1}{\pi\,m_\rho^2}\,\int\limits_0^{s_0}\!\!
ds\,\mbox{Im}\,F_{QCD}^{\pi\gamma^*\gamma^*}(s_1,s)\, 
  \mbox{e}^{\frac{m_\rho^2-s}{M^2}} + \frac{1}{\pi}\,\int\limits_{s_0}^\infty
  \!\!\frac{ds}{s}\,\mbox{Im}\,F_{QCD}^{\pi\gamma^*\gamma^*}(s_1,s).
\end{equation}
This expression is the basic sum rule used for the numerical analysis. 
The calculation of the spectral density of the twist-2 
operator including the ${\mathcal O}(\alpha_S)$ radiative correction 
gives \cite{Schm}
\begin{eqnarray}\label{ammm}
&\frac{1}{\pi}&\mbox{Im}_{s_2}F^{\pi\gamma^*\gamma^*}(s_1,s_2)=
\frac{2\sqrt{2}f_\pi s_2s_1}{\left( s_2 - s_1 \right)^3}
       \Big( 1 +\frac{\alpha_s(\mu)C_F}{12\pi}\cdot \\
&\cdot &\Big( -15 + \pi ^2 - 3\log^2 (-\frac{s_2}{s_1}) \Big)
+ a_2(\mu)\,{A_2}(s_2,s_1) + a_4(\mu)\,{A_4}(s_2,s_1)\nonumber
\Big).
\end{eqnarray}
As usual, the distribution amplitude of twist 2 is expanded 
in Gegenbauer polynomials, keeping only the first three terms:  
 $\varphi_\pi= 6\,u(1-u)\left(1+a_2C_2^{3/2}+a_4 C_4^{3/2}\right).$
The coefficients $A_{2,4}$ in (\ref{ammm})  
are too complicated to be given here. 
They can be found in \cite{Schm}. 

We then combined the twist-2 contribution at NLO 
with the higher twist contributions up to twist 4, 
calculated in \cite{AK}, and  analyzed  
the LCSR for the form factor of the process 
$\gamma\gamma^*\to \pi^0$  numerically.
\begin{figure}[tr]
\psfig{figure = 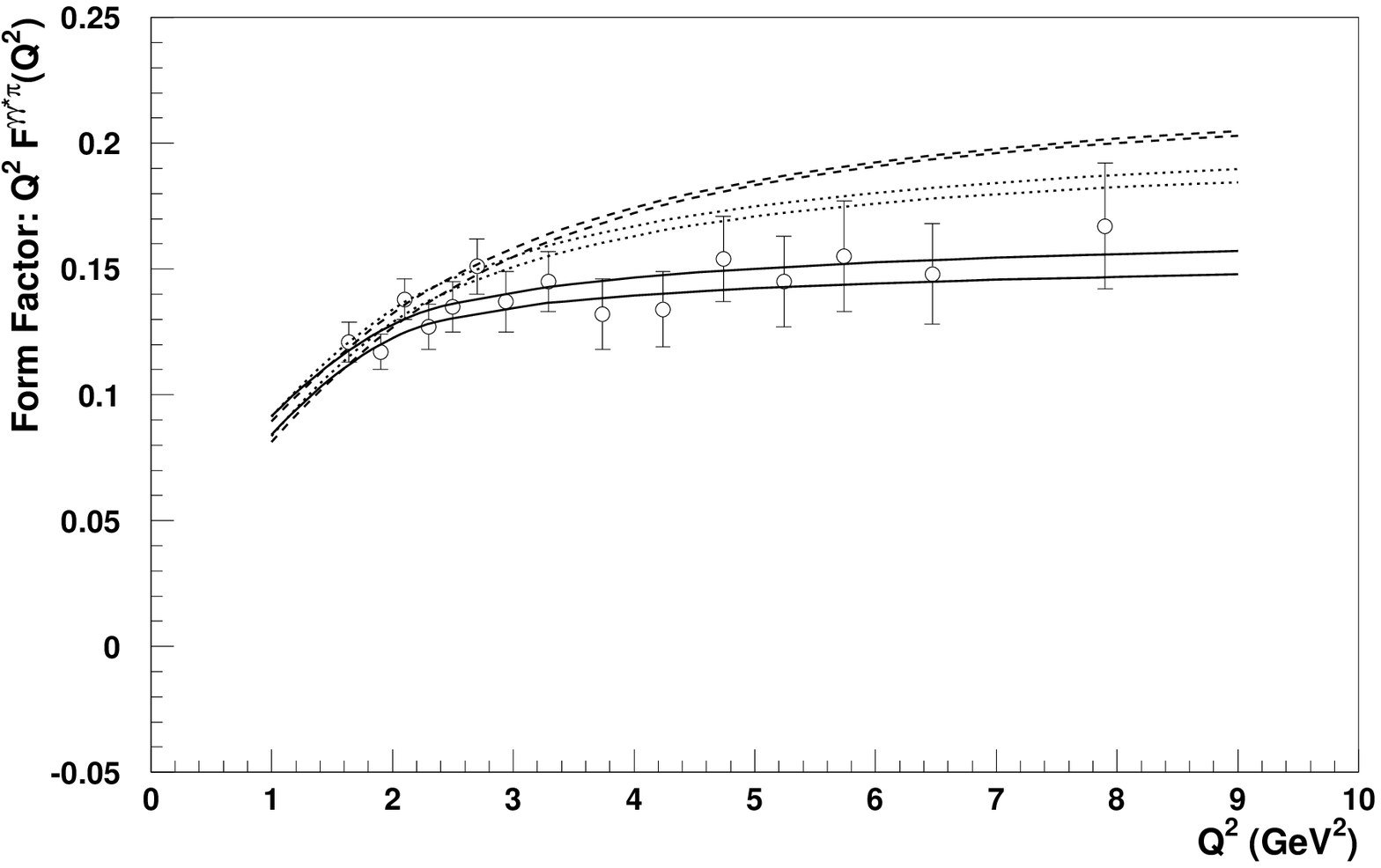,height=1.5in}
\hspace{7mm}
\psfig{figure = 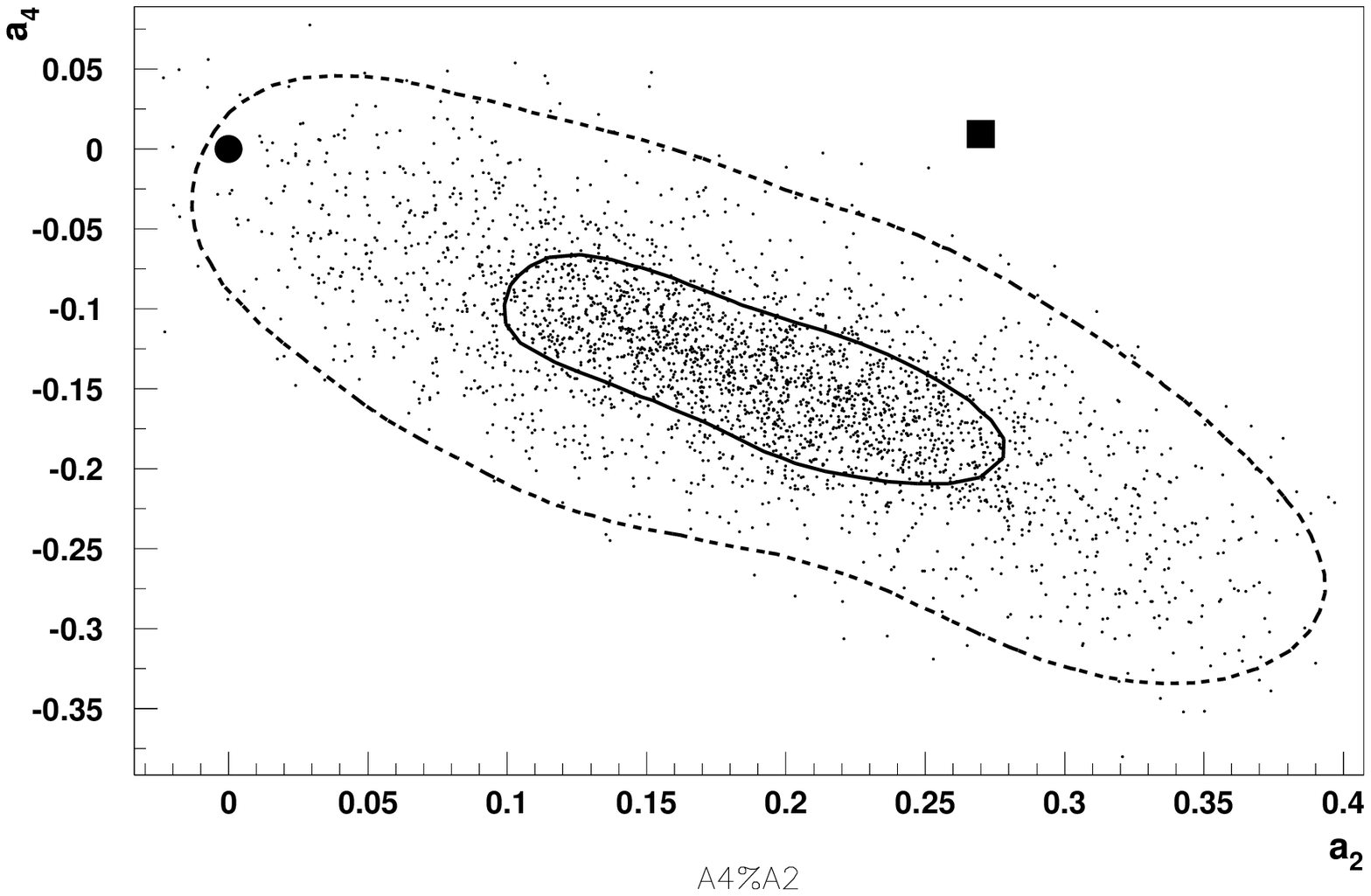,height=1.5in}
\caption{{\bf Left:}
The form factor  $Q^2\,F^{\gamma\gamma^*\pi}(Q^2)$ with 
different distribution amplitudes:
Braun-Filyanov (dashed lines), Chernyak-Zhitnitsky 
(dotted lines) and $\varphi_\pi$ extracted from CLEO data. 
{\bf Right:} Ranges of the coefficients 
$a_2$ and $a_4$ suggested by CLEO data in comparison to the 
Chernyak-Zhitnitsky model (square) and  
 the asymptotic distribution amplitude (circle). 
\label{pion1}}
\end{figure}
Details of the analysis  are given in \cite{Schm}.
 The coefficients $a_2$ and $a_4$ of the  
twist-2 distribution amplitude can be determined    
by comparing the sum rule (\ref{E:srggpi}) with CLEO data \cite{CLEO}. 
 We find that the deviation of the pion distribution amplitude 
from the asymptotic form is small. 
 More definitely, putting $a_4=0$,  we get\cite{Schm} 
\begin{eqnarray}\label{pionres}
a_2(\mu) = 0.12 \pm 0.03\quad\mbox{at}\quad \mu=2.4\quad\mbox{GeV}. 
\end{eqnarray} 
 This result agrees well with a recent 
analysis of the electromagnetic pion form factor \cite{BKM}. 
 Fig. \ref{pion1} shows the form factor  $Q^2\,F^{\gamma\gamma^*\pi}(Q^2)$ 
calculated with different distribution amplitudes: 
Braun-Filyanov\cite{BF} (dashed lines), Chernyak-Zhitnitsky\cite{CZ}   
(dotted lines) and (\ref{pionres}). 
 In principle, one can also extract the coefficient $a_4$.
Unfortunately, the present data is not good enough   
to fix the values of $a_2$ and $a_4$ simultaneously. 
  The ranges of $a_2$ and $a_4$ favored by CLEO data are shown 
in Fig. \ref{pion1}. Obviously, these are in qualitative agreement 
with (\ref{pionres}) and also with the results derived in   
\cite{MikhRad,BakMikh,Belyaev,BKM}, where 
it has also  been claimed that the pion distribution 
amplitude is very close to the asymptotic form.  

\section{Coupling constants $g_{B^*B\pi}$ and $g_{D^*D\pi}$} 
The hadronic $B^*B\pi$ coupling is defined by the 
on-shell matrix element
\be
\langle \bar{B}^{*0}(p)~\pi^-(q)\mid B^-(p+q)\rangle =
-g_{B^*B\pi}(q \cdot\epsilon )~,
\label{def}
\ee
where the meson four-momenta are given in brackets and $\epsilon_\mu$ is the 
polarization vector of the $B^*$. An analogous definition holds
for the $D^*D\pi$ coupling. These couplings play an important role
in $B$ and $D$ physics. For example, they  determine the magnitude
of the weak $B\to\pi$  and $D\to\pi$ form factors 
at zero pion recoil.
Moreover, the coupling  constant $g_{D^*D\pi}$ is directly related 
to the decay width of $D^*\to D \pi$. 
The decay $B^*\to B\pi$ is kinematically forbidden. 
Theoretically, the 
$B^*B\pi$ and $D^*D\pi$ couplings have been 
studied using a variety of 
methods\footnote{An overview is given, e.g., in Tab. 1 of ref. \cite{BBKR}.}. 
Among these, QCD light-cone sum rules (LCSR) have proved particularly 
powerful.  The LCSR calculations of $g_{B^*B\pi}$ and $g_{D^*D\pi}$ 
including perturbative QCD effects in LO and NLO   
were reported in \cite{BBKR,Khodjamirian:1999hb}.
The final LCSR reads
\begin{eqnarray}
f_Bf_{B^*}g_{B^*B\pi}=\frac{m_b^2f_\pi}{m_B^2m_{B^*}}
e^{\frac{m_{B}^2+m_{B^*}^2}{2M^2}}
\Bigg[ M^2 \left(e^{-\frac{m_b^2}{M^2}} - 
e^{-\frac{s_0^B}{M^2}}\right)\varphi_\pi(1/2,\mu)
\nonumber
\\
+\frac{\alpha_s C_F}{4 \pi}
\int\limits_{2m_b^2}^{2s_0^B} 
f \left( \frac{s}{m_b^2}-2 \right)
e^{-\frac{s}{2M^2}}ds  
+ F^{(3,4)}(M^2,m_b^2,s_0^B,\mu)\Bigg]~ 
\label{finalsr}
\end{eqnarray}
with
\begin{eqnarray}
f(x) & = & \frac{\pi^2}{4} +
3 \ln\left(\frac{x}{2}\right) \ln\left( 1+\frac{x}{2}\right ) 
- \frac{3(3 x^3 + 22 x^2 + 40 x + 24)}{2(2+x)^3} 
\ln\left(\frac{x}{2}\right) 
\nonumber
\\ 
& & + 6 \mbox{Li}_2\left(-\frac{x}{2}\right) - 3 \mbox{Li}_2(-x) - 
3 \mbox{Li}_2(-x-1) 
-3 \ln(1+x) \ln(2+x) 
\nonumber
\\
& & -\frac{3(3 x^2 + 20 x + 20)}{4(2+x)^2}
+ \frac{6x (1+x) \ln(1+x)}{(2+x)^3}~. 
\end{eqnarray}
In (\ref{finalsr}), we have added  the contributions $F^{(3,4)}$ from 
the pion distribution amplitudes of twist 3 and 4, which can 
be found in \cite{BBKR}.
\begin{figure}[tr]
\psfig{figure = 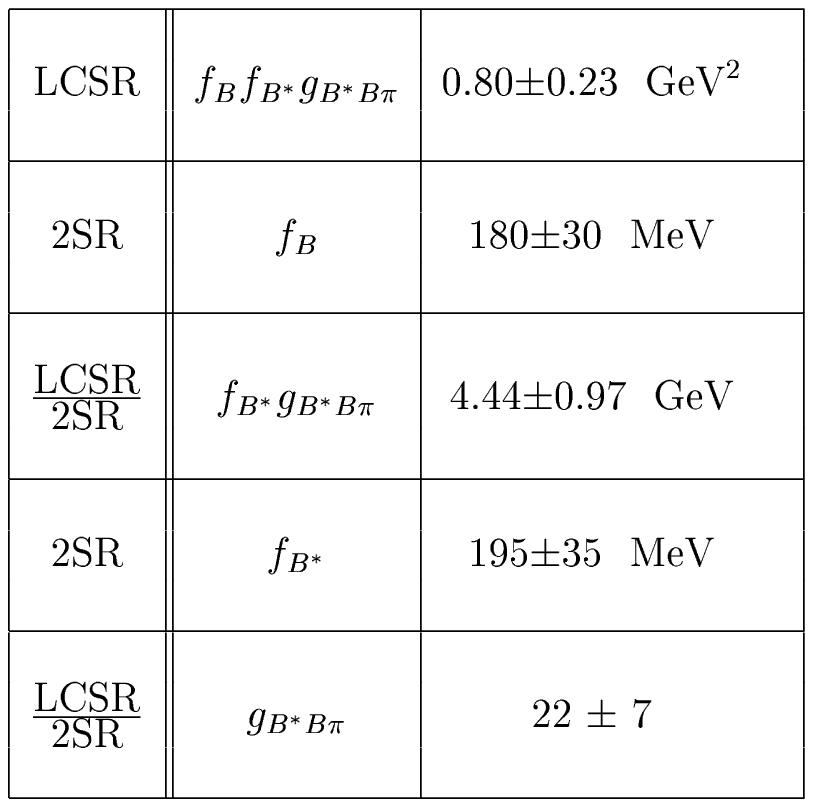,height=2in}
\hspace{10mm}
\psfig{figure = 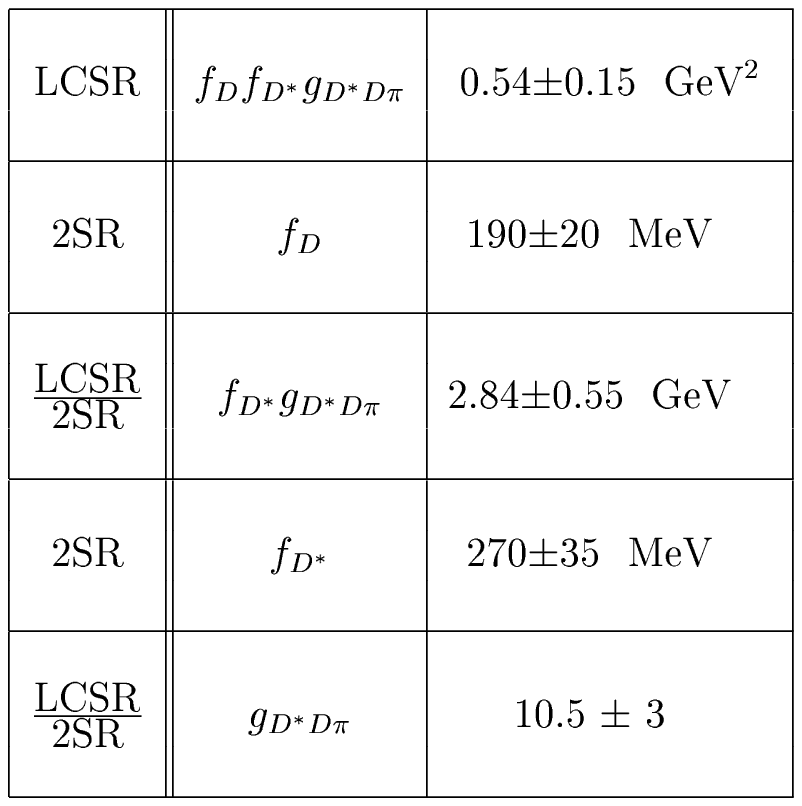,height=2in}
\caption{{\bf Left table:} Sum rule predictions for
$B$ and $B^*$ mesons. {\bf Right table:} Sum rule predictions for
$D$ and $D^*$ mesons. \label{tables}}
\end{figure}
For the $b$-quark mass and the corresponding continuum threshold 
we use $m_b=4.7\pm 0.1$ GeV and  $s_0^B=35\mp 2 $GeV$^2$ respectively. 
The  running coupling constant is taken in the two-loop approximation
with $N_f=4$ and  $\Lambda^{(4)}_{\overline{MS}}=315$ MeV 
corresponding to $\alpha_s(m_Z)=0.118$ \cite{PDG}. 
In the charm case, the corresponding parameters are  given by 
$m_c=1.3\pm 0.1$ GeV, $s_0^D = 6$ GeV$^2$, 
and $\Lambda^{(3)}_{\overline{MS}}=380$ MeV  
\footnote{The meson masses are (in GeV) $m_B=5.279, m_{B^*}=5.325, 
m_D=1.87, m_{D^*}=2.01,$ and $f_{\pi}=132$ MeV. }.  
Finally, for the pion distribution amplitude $\varphi_\pi(u, \mu)$ 
at $u=0.5$ and  $\mu= 2.4$ GeV we have 
$\varphi_\pi(1/2, \mu)=1.23 $ \cite{Schm}. 
 The decay constants and resulting coupling constants  are 
summarized in the two tables in Fig. \ref{tables}. 
We will make use of them in the next section. 
Here we just mention that from $g_{D^*D\pi}$ given in Table 2 
one obtains  
$\Gamma(D^{*+}\rightarrow D^0 \pi^+)=23\pm 13 \mbox{keV}$.  
The current experimental limit \cite{PDG} 
$\Gamma (D^{*+}\to D^0\pi^+) < 89~\mbox{keV}$
is still too high to challenge the theoretical prediction.
\section{ The scalar form factor $f^0$}
In general, the hadronic matrix element of the $B\to \pi$ 
transition is determined by two independent form factors, 
$f^+$ and $f^-$: 
\begin{eqnarray}\label{TransitionAmplitude}
\langle\pi (q)|\bar{u}\gamma_\mu b|B(p+q)\rangle = 2f^+ (p^2) q_\mu + 
\left(f^+(p^2)+f^-(p^2)\right)p_\mu ,
\end{eqnarray}
where $(p+q)$ and $q$ denote the initial and final state four-momenta
and  $\bar u\gamma_\mu b$ is the relevant weak current. 
The form factor $f^{0}$ is usually defined through the matrix element
\begin{eqnarray}\label{F0}
p^{\mu}\langle\pi (q)|\bar{u}\gamma_\mu b |B(p+q)\rangle 
= f^0(p^2)(m_B^2-m_{\pi}^2),
\end{eqnarray}
yielding together with (\ref{TransitionAmplitude}) 
$f^0=f^++\frac{p^2}{m_B^2 - m_\pi^2}f^-.$
In order to determine $f^0$ from sum rules it is advantageous to
consider $f^+$ and $f^++f^-$. The sum rule for $f^+$ has been 
analysed in \cite{K2000,KRWY2}. 
 The sum rule for the sum of form factors is given by 
\begin{eqnarray}\label{SumRule}
f^++f^-=-\frac{m_bf_\pi}{\pi \, m_B^2\,f_B}
\int\limits_{m_b^2}^{s_0}\!\!ds \int\limits_0^1\!\! du 
\,\exp\left(-\frac{s-m_B^2}{M^2}\right)\,\varphi_\pi (u)\,
\mbox{Im}\,\tilde T_{QCD},
\end{eqnarray} 
where $M$ again denotes the Borel mass.   
The expression of the hard amplitude $\tilde T_{QCD}(p^2,s,u,\mu)$
 in  LO and NLO can be found in \cite{KRW} and in\cite{Fzero},  
respectively. 
The leading twist-2 contribution to the imaginary  
part of the hard amplitude is given by \cite{Fzero} 
\begin{figure}[tr]
\psfig{figure = 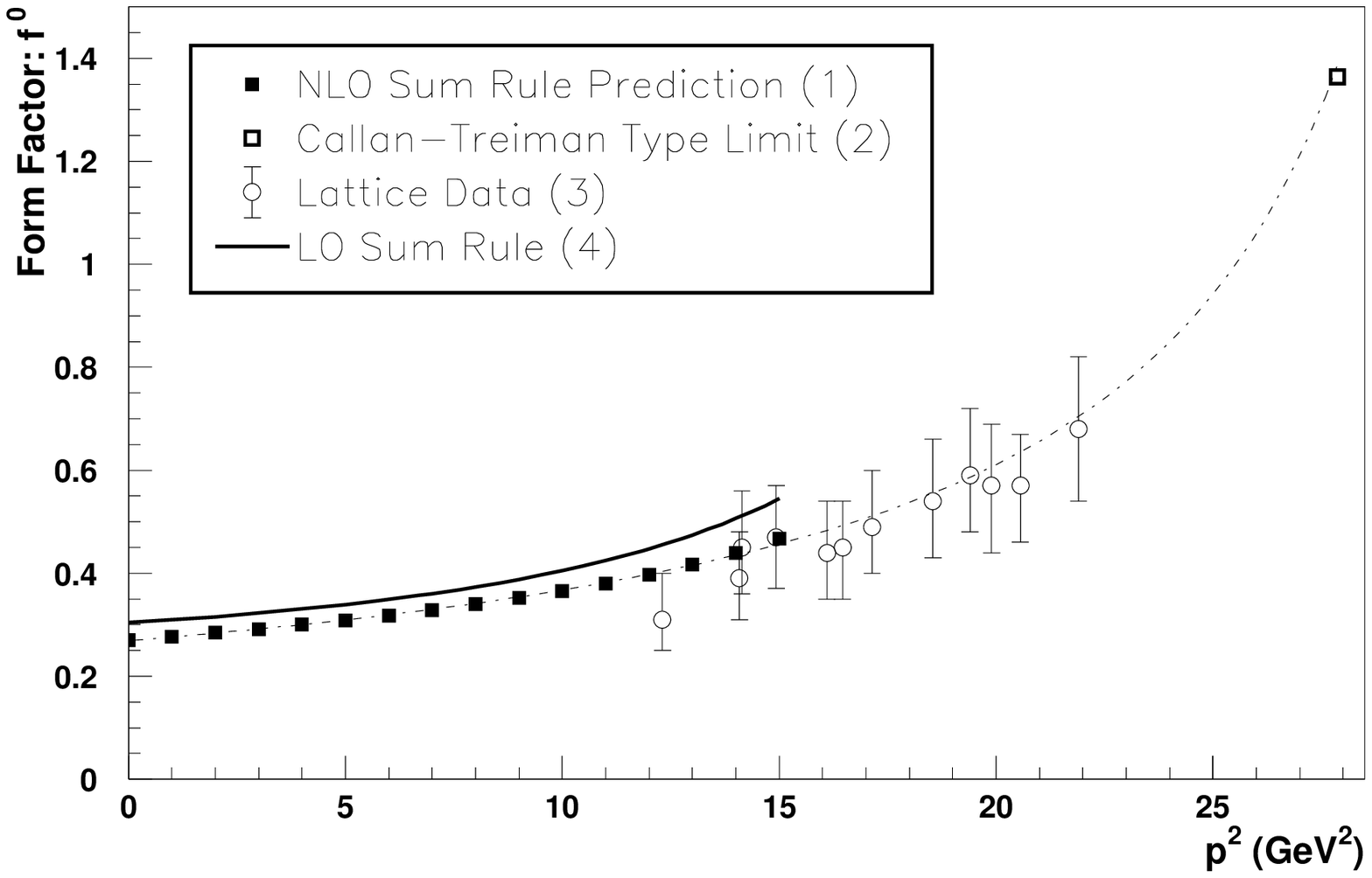,height=1.8in}
\hspace{5mm}
\psfig{figure = 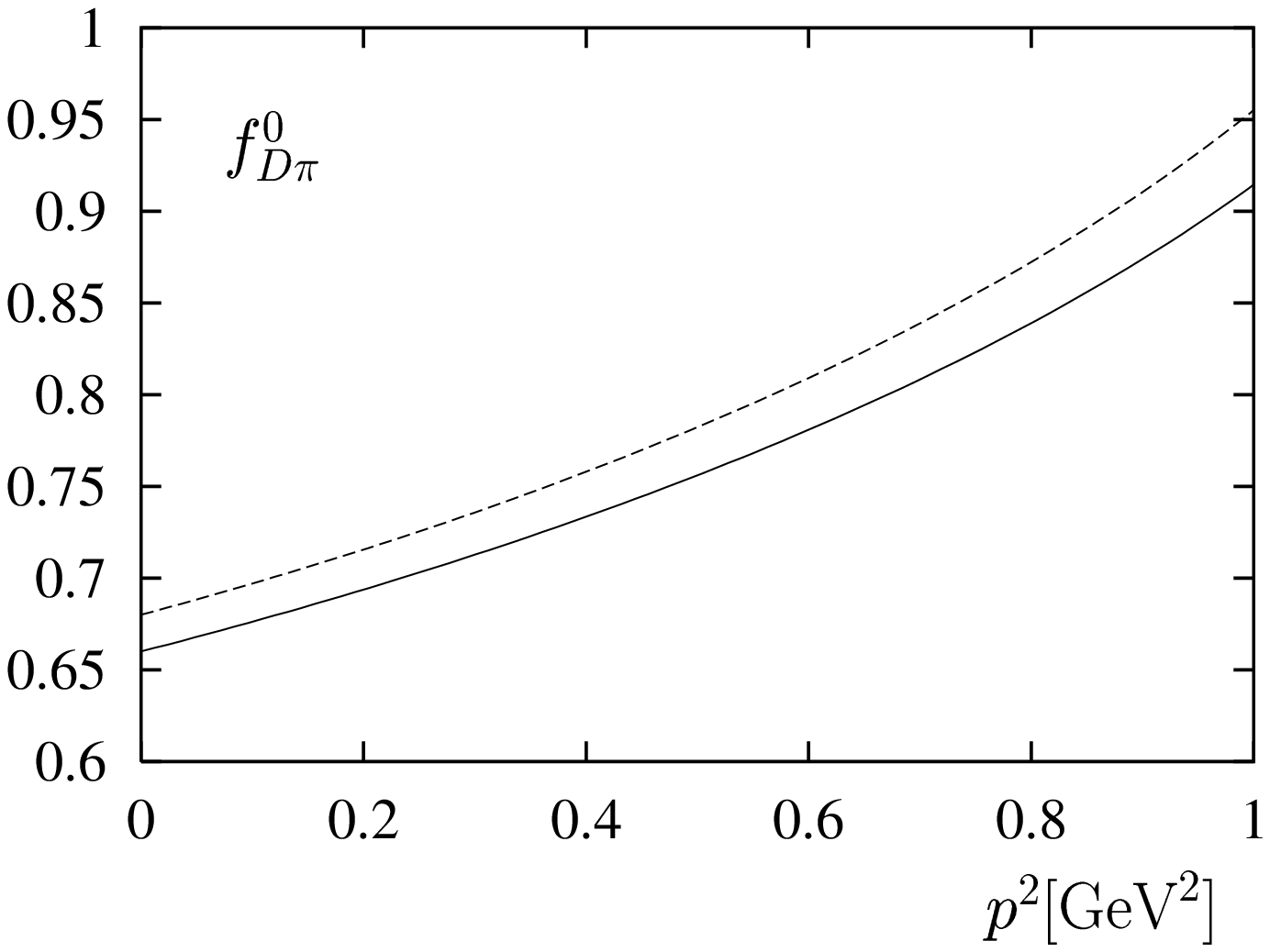,height=1.4in}
\caption{ {\bf Left:}  
NLO LCSR prediction for $f_{B\pi}^0$ (solid squares) 
and fit to the later and the PCAC constraint shown 
by the empty square (dashed line). Also shown are the 
LO LCSR result (solid line) and the UKQCD lattice data (empty circles).
{\bf Right:} Light-cone sum rule results for the form factor 
$f^0_{D\pi}$ in NLO (solid line) and  LO (dashed line). \label{fzero}}
\end{figure}
\begin{eqnarray}\label{ImaginaryPart}
&&\frac{1}{\pi}\mbox{Im}\tilde T_{QCD}(s_1,s_2,u,\mu) =
\Big(\frac{C_F\alpha_s(\mu)}{2\pi}\Big)
\Theta(s_2 - m_b^2) \frac{m_b}{s_2 - s_1}
\nonumber\\ 
&&\Bigg\{ \Theta(u - u_0)
\Bigg[-\frac{(1-u)(u-u_0)(s_2-s_1)^2}{2 u \rho^2} - \frac{1}{u(1-u)}
\left(\frac{m^2}{\rho}-1\right)\Bigg]\\ \nonumber
&+&\delta(u-u_0) \frac{1}{2 u} \Bigg[\frac{(s_1-m_b^2)^2}{s_1^2}
\ln\left(1-\frac{s_1}{m_b^2} \right)
+\frac{m_b^2}{s_1} - 1 \Bigg] 
-\frac{1}{1-u}\left(1-\frac{m_b^2}{s_2}\right)
\Bigg\}
\end{eqnarray}
with $u_0 = \frac{m_b^2 - s_1}{s_2 - s_1}$.
In Fig. \ref{fzero}, the form factor $f^0_{B\pi}$ is plotted together 
with the UKQCD lattice results \cite{Flynn}. 
We see that the radiative contributions improve the  agreement between
the  lattice and  the LCSR calculations.  
Also shown in Fig. 3 are the LCSR results for the form factor 
$f^0_{D\pi}$.
 We note that $f^0_{D\pi}(0)=0.66$. 
\section{New method of  calculating  $f^+$ }
In this section we review a new method suggested in \cite{WY} 
for calculating heavy-to-light form factors.
 The method is based on first principles. It is  
an extension of LCSR, but it has a much wider  
range of applicability, including the intermediate momentum region, 
where most of the lattice results are located,  
and even the region near zero recoil.   
 The main idea is to use the operator product expansion with  
a combination of double and single dispersion relations. 
The resulting new sum rule has a term  which   
 corresponds to the ground-state, as well as contributions  
which account for all possible physical intermediate states. 
 We start from the usual correlation function
\bq
F_\mu(p,q)& = & i\int dx e^{ip\cdot x}
\langle \pi(q)|T\{\bar u(x)\gamma_\mu b(x),
m_b\bar b(0)i\gamma_5 d(0)\}|0 \rangle \nonumber \\
&=& F(p^2,(p+q)^2) q_\mu + \tilde{F}(p^2,(p+q)^2) p_\mu, 
\eq
 focusing on the invariant amplitude
$F(p^2,(p+q)^2)$. In the following,  we use the definitions 
\bq
\sigma(p^2,s_2)  =  \frac{1}{\pi} \; \mbox{Im}_{s_2} 
\; F(p^2,s_2), \quad 
\rho(s_1,s_2) =  \frac{1}{\pi^2} \; \mbox{Im}_{s_1} \; 
\mbox{Im}_{s_2} \; F(s_1,s_2). 
\eq
The standard sum rule for the form factor $f^+(p^2)$ is obtained 
 by writing a single dispersion relation 
for $F(p^2,(p+q)^2)$ in the $(p+q)^2$-channel, inserting the
hadronic representation for $\sigma(p^2,s_2)$ and Borelizing in $(p+q)^2$:
\bq
\label{eqa1}
{\cal B}_{(p+q)^2} F & = & {\cal B}_{(p+q)^2} 
\left( \frac{2 m_B^2 f_B f^+(p^2)}{m_B^2-(p+q)^2} + 
\int\limits_{s_2>s_0} ds_2 \frac{\sigma^{hadr}(p^2,s_2)}{s_2-(p+q)^2} 
\right).
\eq
Note that any subtraction terms which 
might appear vanish after Borelization.
 Similarly, the standard light-cone sum rule for the coupling 
$g_{B^\ast B \pi}$ is obtained from a double dispersion
relation:
\bq
\label{eqa2}
{\cal B}_{p^2} {\cal B}_{(p+q)^2} F & = &  
{\cal B}_{p^2} {\cal B}_{(p+q)^2} \left( \frac{m_B^2 m_{B^\ast} f_B f_{B^\ast} g_{B^\ast B \pi}}
{(p^2 - m_{B^\ast}^2)( (p+q)^2 - m_B^2)} \right. \nonumber \\
& & \left. + \int\limits_{\Sigma} ds_1 ds_2 \frac{\rho^{hadr}(s_1,s_2)}{(s_1-p^2)(s_2-(p+q)^2)} \right),
\eq
where $\Sigma$ denotes the integration region 
defined by $s_1>s_0$, $s_2>m_b^2$ and $s_1>m_b^2$, $s_2>s_0$.  

In contrast to the above procedure we suggest to use a dispersion relation 
for $\sigma(p^2,s_2)/(p^2)^l$ in the $p^2$-channel 
(with $l$ being an integer):
\bq
\label{eqa3}
\hspace{-6mm}\sigma(p^2,s_2) & = & - \frac{1}{(l-1)!} \left(p^2\right)^l \frac{d^{l-1}}{ds_1^{l-1}}
\left. \frac{\sigma(s_1,s_2)}{s_1-p^2} \right|_{s_1=0} 
+ \int\limits_{s_1>m_b^2} ds_1 \frac{(p^2)^l}{s_1^l} 
\frac{\rho(s_1,s_2)}{s_1 -p^2},  
\eq
and to replace $\sigma(p^2,s_2)$ in (\ref{eqa1})  
by the r.h.s of (\ref{eqa3})
\footnote{By choosing $l$ large 
enough the dispersion relation (\ref{eqa3}) 
will be convergent.}.
Then, writing a double dispersion 
relation for $F(p^2,(p+q)^2)/(p^2)^l$ and 
comparing it with the previous result, we obtain the sum rule
\bq
\label{eqa6}
f^+(p^2) & = & \frac{1}{2} \frac{(p^2)^l}{(m_{B^\ast}^2)^l} \frac{f_{B^\ast} g_{B^\ast B \pi}}
{m_{B^\ast} \left( 1 -\frac{p^2}{m_{B^\ast}^2} \right)} 
- \frac{1}{(l-1)!} \left( p^2 \right)^l \left. \frac{d^{l-1}}{ds_1^{l-1}} 
\frac{f^+(s_1)}{s_1-p^2} \right|_{s_1=0} \nonumber \\
& & + \frac{1}{2 m_B^2 f_B} \int\limits_{\Sigma'} ds_1 ds_2 
\frac{(p^2)^l}{s_1^l} \frac{\rho(s_1,s_2)}{s_1-p^2} e^{- \frac{s_2-m_B^2}{M^2}},
\eq
where the integration region $\Sigma'$ is defined by $s_1>s_0$ and 
$m_b^2 < s_2 < s_0$. This sum rule is valid in the whole 
kinematical range of $p^2$.  As input we need the 
first $(l-1)$ terms of the Taylor expansion of $f^+(p^2)$ 
around $p^2=0$. These parameters can be obtained 
numerically  from the standard sum rule for $f^+(p^2)$:
\bq
\label{eqa1a}
f^+(p^2) & = & \frac{1}{2m_B^2f_B} \int\limits_{m_b^2}^{s_0} \sigma^{QCD}(p^2,s_
2)
e^{-\frac{s_2-m_B^2}{M^2}}
\eq
following from (\ref{eqa1}). 
We further need the residue at the pole $p^2=m_{B^\ast}^2$, 
which can be obtained from the sum rule (\ref{eqa2}), 
as discussed in the previous section (see (8)).  

 The case $l=0$ has a very transparent physical meaning. 
The first term represents the contribution of the ground state 
resonance with mass $m_{B^*}$, while the second term corresponds 
to the contributions of all other physical states in this channel.
 As shown explicitly in\cite{WY} for twist 2, 3 and 4,  
the last term is of  $O(\alpha_s)$ only.   
 This provides an explanation of the empirical fact  
that the single pole model describes many form factors quite well.   
In addition, we are now able to quantify the deviation from the pole model,
in a model-independent way, applying QCD and light-cone OPE.   
 It should be noted that the parameter $l$ plays a 
similar role as the Borel parameter $M^2$. 
There is a lower limit on $l$ such that the dispersion relation  
 (\ref{eqa3}) converges. 
Going to higher values of $l$ will improve the convergence of the 
dispersion relations
and will suppress higher resonances in the $B^\ast$-channel.
But there is also an upper limit on $l$.  
 The higher the value of $l$, the more derivatives
of $f^+(p^2)$ at $p^2=0$ enter. At some point, one starts probing 
the region $p^2 > m_b^2 - 2 \chi m_b$, where the standard sum  
rule (\ref{eqa1a}) breaks down.  
\begin{figure}[tr]
\psfig{figure =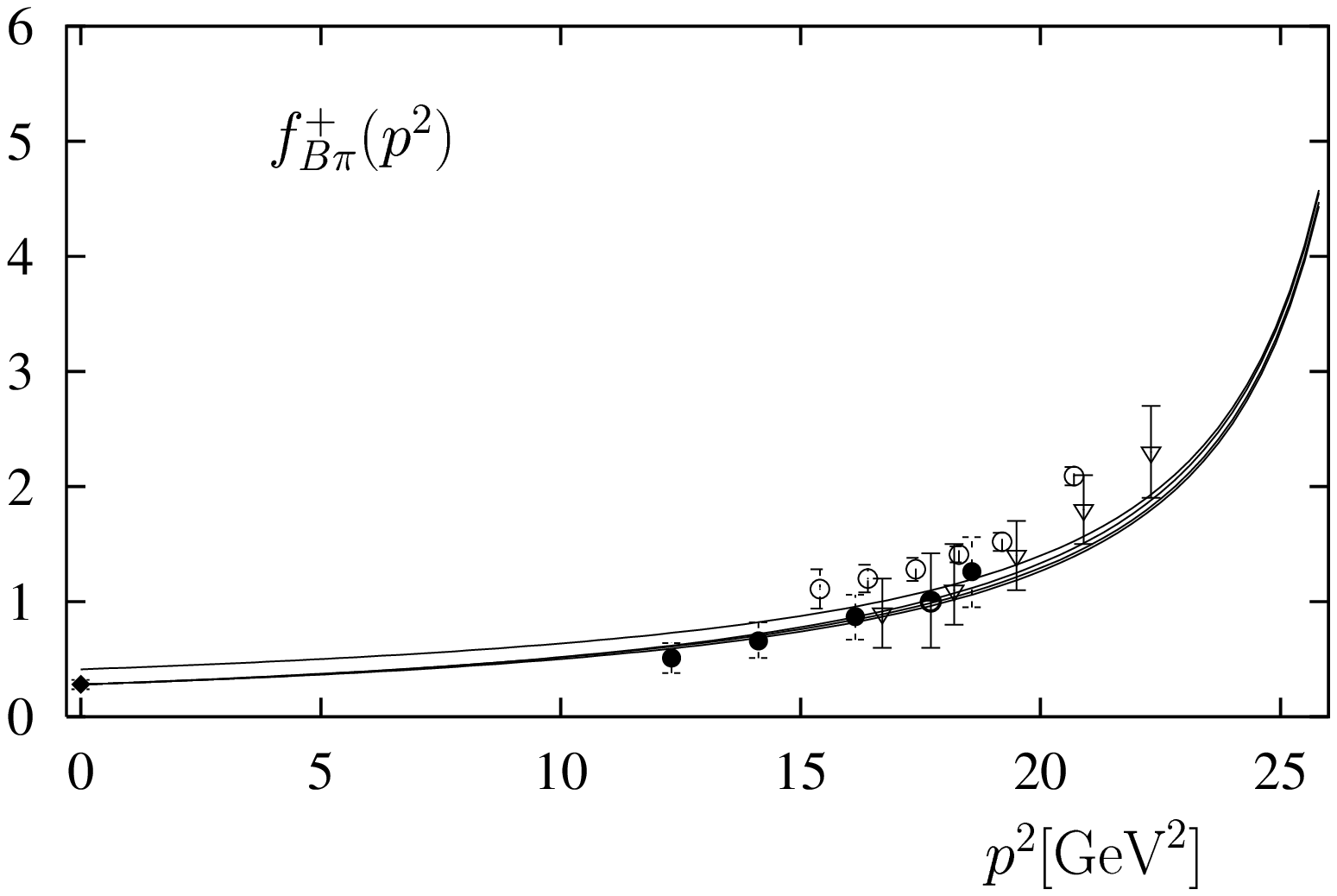,height=1.5in}
\psfig{figure=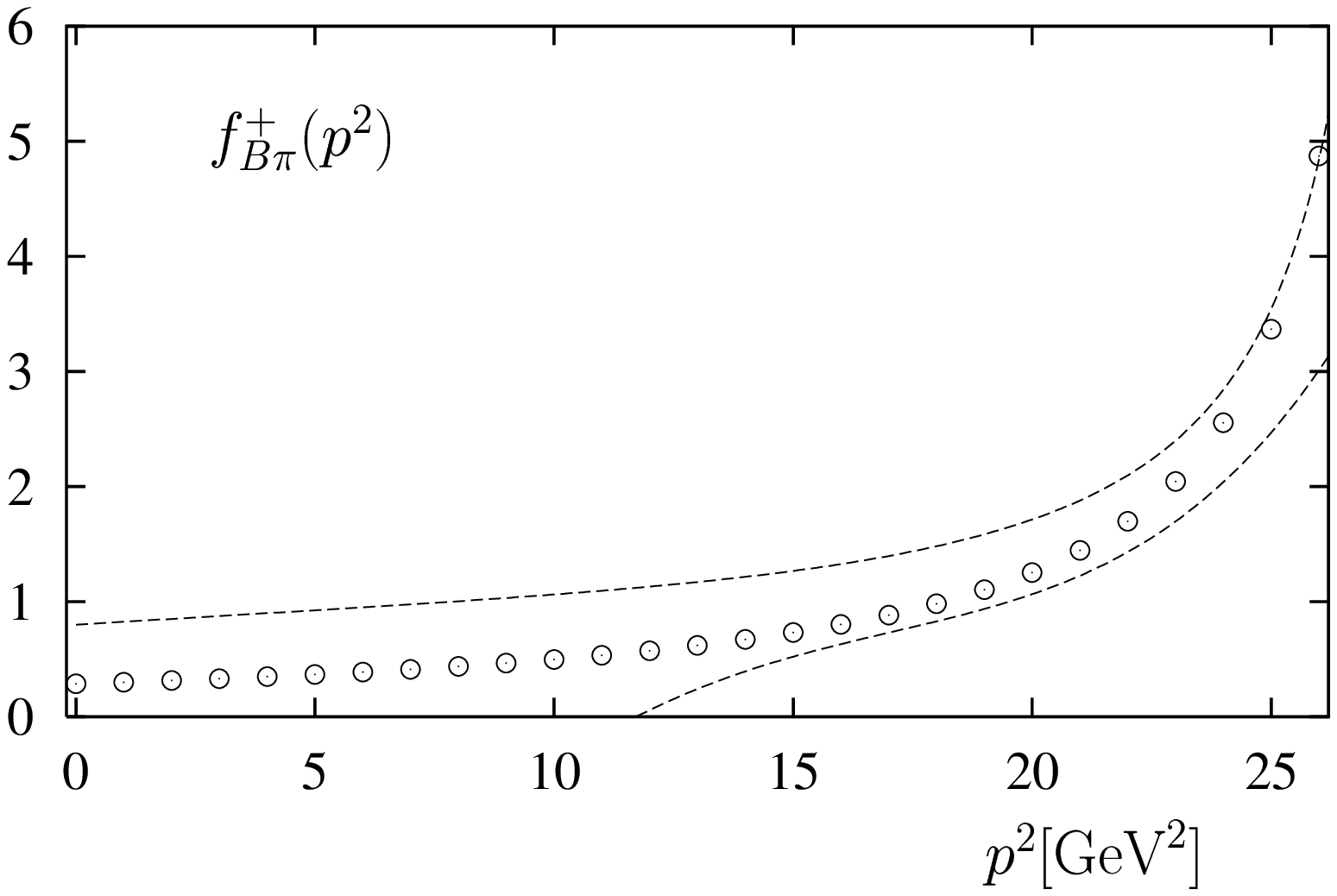,height=1.5in}
\caption{ {\bf Left:} 
The LCSR prediction for the $B\to\pi$ 
form factor at $l=0,1,2,3$ in comparison to lattice results. 
 The lattice results come from FNAL (full circles),
UKQCD (triangles), APE (full square), 
JLQCD (open circles), and ELC  (semi-full
circle).  {\bf Right:} 
The LCSR prediction on the form factor $f_{B\pi}^+$ (circles)
in comparison to the constraint (dashed) derived by
Boyd and Rothstein.\label{fp1} } 
\end{figure}
Details on the numerical analysis of the new sum rule 
can be found in \cite{WY}. 
 This analysis nicely supports the qualitative results obtained in
\cite{K2000}. Using the convenient parameterization\cite{BK}  
\be\label{paramB}
f_{B\pi}^+(p^2)=\frac{f_{B\pi}^+(0)}{(1-p^2/m_{B^*}^2)(1-
\alpha_{B\pi}p^2/m_{B^*}^2)} ~, 
\ee
and $f^+_{B\pi}(0) = 0.28 \pm 0.05$\cite{K2000,KRWY2}, 
we get $\alpha_{B\pi} =  0.4 \pm 0.04~$ in  remarkable agreement with 
$\alpha_{B\pi} =  0.32 \pm^{0.21}_{0.07}$ derived in \cite{K2000}.  
 Fig. \ref{fp1} shows a comparison of (\ref{paramB}) with
recent lattice results \cite{Flynn,lattrev,UKQCD,APE,JLQCD}.
The agreement within uncertainties is very satisfactory.
Finally, the LCSR prediction also obeys the constraints
derived from sum rules for the inclusive
semileptonic decay width in the heavy quark limit \cite{Boyd}.
This is also demonstrated in Fig. \ref{fp1}.

The above results on $f^+_{B \pi}$ can be used 
to calculate the width of the semileptonic decay  
 $B\to \pi \bar{l}\nu_l$ with $l=e,\mu$. 
For the integrated width, one obtains\cite{K2000} 
\be
\Gamma 
= \frac{G^2|V_{ub}|^2}{24\pi^3} \int  dp^2 
(E_\pi^2-m_\pi^2)^{3/2}\left[f^+_{B\pi}(p^2)\right]^2
= (7.3 \pm 2.5) ~|V_{ub}|^2~\mbox{ps}^{-1}~.
\label{elmu}
\ee
Experimentally, combining the branching ratio 
$BR(B^0\to\pi^-l^+\nu_l) = (1.8 \pm 0.6)\cdot 10^{-4}$
with the $B^0$ lifetime
$\tau_{B^0}=1.54 \pm 0.03$ ps \cite{PDG},   
one gets 
$
\Gamma(B^0\rightarrow \pi ^- l^+ \nu_l) =
(1.17 \pm 0.39)\cdot10^{-4}~\mbox{ps}^{-1}~.
$
From that and (\ref{elmu})
one can then determine the quark mixing parameter $|V_{ub}|$.
The result is
\be
|V_{ub}|=(4.0 \pm 0.7 \pm 0.7)\cdot 10^{-3}
\label{result}
\ee
with the experimental error 
and theoretical uncertainty given in this order.
For the $D\to\pi$ transition and using (\ref{paramB}) analogously one obtains \cite{K2000} 
$\alpha_{D\pi} = 0.01^{+0.11}_{-0.07}~$  and  
$f^+_{D\pi}(0) = 0.65 \pm 0.11,$   
which nicely agrees with lattice estimates,  
 for example, the world average \cite{lattrev} 
$f^+_{D\pi}(0)= 0.65 \pm 0.10 ~,$
or the most recent APE result \cite{APE}, 
$f^+_{D\pi}(0)= 0.64 \pm 0.05 ^{+.00}_{-.07}$.
For more details one should consult\cite{K2000}.
\section{Heavy baryons}
The study of heavy baryons such as 
$\Lambda_b,\Lambda_c$ and $\Sigma_b, \Sigma_c$, 
is more complicated. Two and three point QCD sum rules  
have been investigated in 
\cite{Grozin1,Groote1,Groote2,Groote3,Olegrev}, and have been applied 
to the heavy-to-light baryon transitions\cite{Huang1,Huang2,Carvalho}. 
However, we are not aware of applications of LCSR to heavy-to-light 
baryon transitions. 
In the following we collect the results available at present (see also \cite{Olegrev}).  
One has estimated the binding energies,
$\bar \Lambda =M - m_Q$, 
 of the ground state baryons 
and the residues of the baryonic currents, 
$\langle B|J_B|0 \rangle = F_B u_B$,  at NLO. 
The results are\cite{Grozin1,Groote3,Groote1,Groote2} 
\be
\bar \Lambda (\Lambda_Q ) = 0.77 \pm 0.05 \mbox{GeV} \quad \mbox{and} \quad 
|F_{\Lambda_Q}| = 0.027 \pm 0.001 \mbox{GeV}^3,
\ee
\be
\bar \Lambda(\Sigma_Q) = 0.94 \pm 0.05 \mbox{GeV}\quad\mbox{and}\quad
|F_{\Sigma_Q}|=0.038 \pm 0.003\mbox{GeV}^3. 
\ee
\newpage
\hspace{-6mm}Using the experimental value for the mass of the $\Lambda_b$ baryon,
$ m(\Lambda_b)=5.642 \pm 0.05 $ GeV \cite{PDG} one  finds 
for the pole mass of the $b$ quark:  
$ m_{b}=4.88 \pm 0.1$, and for  the related 
$\overline {MS}$ mass:  $\bar m( \bar m ) = 4.25 \pm 0.1 \mbox{GeV}.$ 
These values are in good agreement with the  
mass estimates in the meson case\cite{Voloshin,BenekeB}.
 Coupling constants have been derived from 
sum rules in the external axial field \cite{Grozin3} 
with the  result: 
$g_{\Sigma^*\Sigma\pi} = 0.83 \pm 0.3$ and  
$g_{\Sigma^*\Lambda\pi}= 0.58 \pm 0.2 $.
 The semileptonic transition $\Lambda_b\to\Lambda_c$ has been 
studied\cite{Grozin2} using sum rule techniques. 
The matrix elements of this weak transition are determined 
by the Isgur-Wise function
\be
\langle \Lambda_b|\bar c \Gamma b |\Lambda_c \rangle = \xi(w) \bar u_c \Gamma u_b
\ee
where $w=v_b \cdot v_c$ and $\Gamma$ is the Dirac matrix. 
The baryonic Isgur-Wise function has been estimated by QCD 
sum rules in \cite{Grozin2}. The slope 
of this function at 
$w=1$ is found to be $\rho^2=-1.15\pm 0.2$ and the shape 
is fitted very well by $\xi (w) 
=\frac{2}{1+w} exp\Big((2\rho^2-1)\frac{w-1}{w+1} \Big).$
Taking into account $1/m$ corrections\cite{Dai1}, one obtains the 
width $\Gamma (\Lambda_b\to\Lambda_c e\nu) = 
|\frac{V_{cb}}{0.04}|^2\cdot 6 \cdot 10^{-14}~\mbox{GeV}.$\\

{\bf Conclusion:} We have given a short review of selected 
topics concerning QCD sum rules for heavy hadrons. 
One main development during the past few years 
 has been the NLO improvement. A second important 
development is the ongoing update of the pion wave functions \cite{Schm}. 
 Very recently a new sum rule for $B\to \pi$ has been suggested\cite{WY}. 
 Among the remaining problems, 
we have mentioned the application  of LCSR to baryons 
requiring the knowledge of baryonic distribution amplitudes. \\

{\bf Acknowledgments:} O.Y. acknowledges support from the US 
Department of Energy.
R.R. acknowledges support from the 
 Bundesministerium f\"ur Bildung und Forschung  
under contract number 05HT9WWA9. 

\end{document}